\documentclass[3p,times,procedia]{elsarticle}
\usepackage{nupha_ecrc}

\volume{00}

\firstpage{1}

\journalname{Nuclear Physics A}

\runauth{Pascal Dillenseger for the ALICE Collaboration}

\jid{nupha}

\jnltitlelogo{Nuclear Physics A}

\usepackage{amssymb}

\usepackage{lineno}

\biboptions{sort&compress}

\usepackage[figuresright]{rotating}
\usepackage{subfig,wrapfig}
\usepackage{graphicx}

\begin{document}

\begin{frontmatter}



\dochead{XXVIIth International Conference on Ultrarelativistic nucleus$-$nucleus Collisions\\ (Quark Matter 2018)}

\title{Quarkonium measurements in nucleus$-$nucleus collisions with ALICE}


\author{Pascal Dillenseger for the ALICE Collaboration}

\address{Institut f\"ur Kernphysik - Goethe-Universit\"at, Max-von-Laue-Stra\ss e 1, 60438 Frankfurt am Main - Germany}

\begin{abstract}
	Quarkonia, i.e. bound states of $b\bar{b}$ and $c\bar{c}$ quarks, are powerful observables to study the properties of nuclear matter under extreme conditions. The formation of a Quark-Gluon Plasma (QGP), which is predicted by lattice QCD calculations at high temperatures as reached at LHC energies, has a strong influence on the production and behavior of quarkonia.%
	\\ \indent The latest ALICE results on bottomonium and charmonium production in nucleus$-$nucleus collisions are presented. This includes measurements of the $\Upsilon$(1S) and $\Upsilon$(2S) nuclear modification factors ($R_{\mathrm{AA}}$) at forward rapidity and the J/$\psi$ $R_{\mathrm{AA}}$ and $v_{2}$ as a function of centrality, $p_{\mathrm T}$ and rapidity in Pb$-$Pb collisions at $\sqrt{s_{\mathrm{NN}}}$ = 5.02 TeV. Also, first results from J/$\psi$ measurements in Xe$-$Xe collisions at $\sqrt{s_{\mathrm{NN}}}$ = 5.44 TeV are presented. Further on, the experimental results are compared to various calculations from theoretical models.
\end{abstract}


\end{frontmatter}


\section{Quarkonium production in nucleus$-$nucleus collisions}
\label{sec:Introduction}
The significant enhancement of the J/$\psi$ nuclear modification factor ($R_{\mathrm{AA}}$) at LHC energies compared to SPS and RHIC energies \cite{Adare:2006ns, Adam:2016rdg} indicates an almost leveled competition between suppression and (re)generation of charmonium states in nucleus$-$nucleus collisions in the TeV regime. The $R_{\mathrm{AA}}$ is defined as the ratio of production yields in nucleus$-$nucleus collisions ($\mathrm{d} N_{\mathrm{AA}}$) and pp collisions ($\mathrm{d} N_{\mathrm{pp}}$) scaled with number of binary nucleon-nucleon collisions ($N_{\mathrm{Coll}}$): $R_{\mathrm{AA}} = \frac{ \mathrm{d}N_{\mathrm{AA}} }{ N_{\mathrm{Coll}}~\times~\mathrm{d} N_{\mathrm{pp}} }$. Heavy-quark pairs, i.e. $b\bar{b}$ and $c\bar{c}$, are produced in initial hard processes, thus the quarkonium bound state production is subject to the full evolution of the collision. A suppression of quarkonium states in a Quark-Gluon Plasma (QGP) is expected due to the color-screening effect, which is based on a Debye screening of the color charge \cite{Matsui:1986dk}. In competition with the color-screening effect, where initial $Q\bar{Q}$ pairs are separated and not able to form a bound state, is the (re)combination effect. It is based on the non-zero probability that quasi-free quarks and anti-quarks move close enough in space and momentum to form a quarkonium bound state. The influence on the quarkonium production rates of this effect is strongly coupled to the $Q\bar{Q}$ production cross-section ($\sigma_{Q\bar{Q}}$) \cite{Thews:2000rj, BraunMunzinger:2000px}. Both effects are highly sensitive to the properties of the QGP and should induce diverging behavior on the differential quarkonium production, e.g. momentum distribution and elliptic flow.\\ \indent
In the following, the ALICE results for quarkonium measurements in nucleus$-$nucleus collisions presented during the Quark Matter Conference 2018 are discussed. This includes the J/$\psi$-$R_{\mathrm{AA}}$ measurement in Xe$-$Xe collisions at $\sqrt{s_{\mathrm{NN}}}$ = 5.44 TeV, multi-differential J/$\psi$-yield measurements and differential measurements of the bottomonium $R_{\mathrm{AA}}$ in the forward rapidity range and the elliptic flow of J/$\psi$ at forward and mid-rapidity in Pb$-$Pb collisions at $\sqrt{s_{\mathrm{NN}}}$ = 5.02 TeV.
\section{Experimental results of quarkonium measurements in nucleus$-$nucleus collisions}
\label{sec:Results}
\begin{wrapfigure}{r}{0.475\linewidth}%
	\centering
	\includegraphics[width=0.95\linewidth]{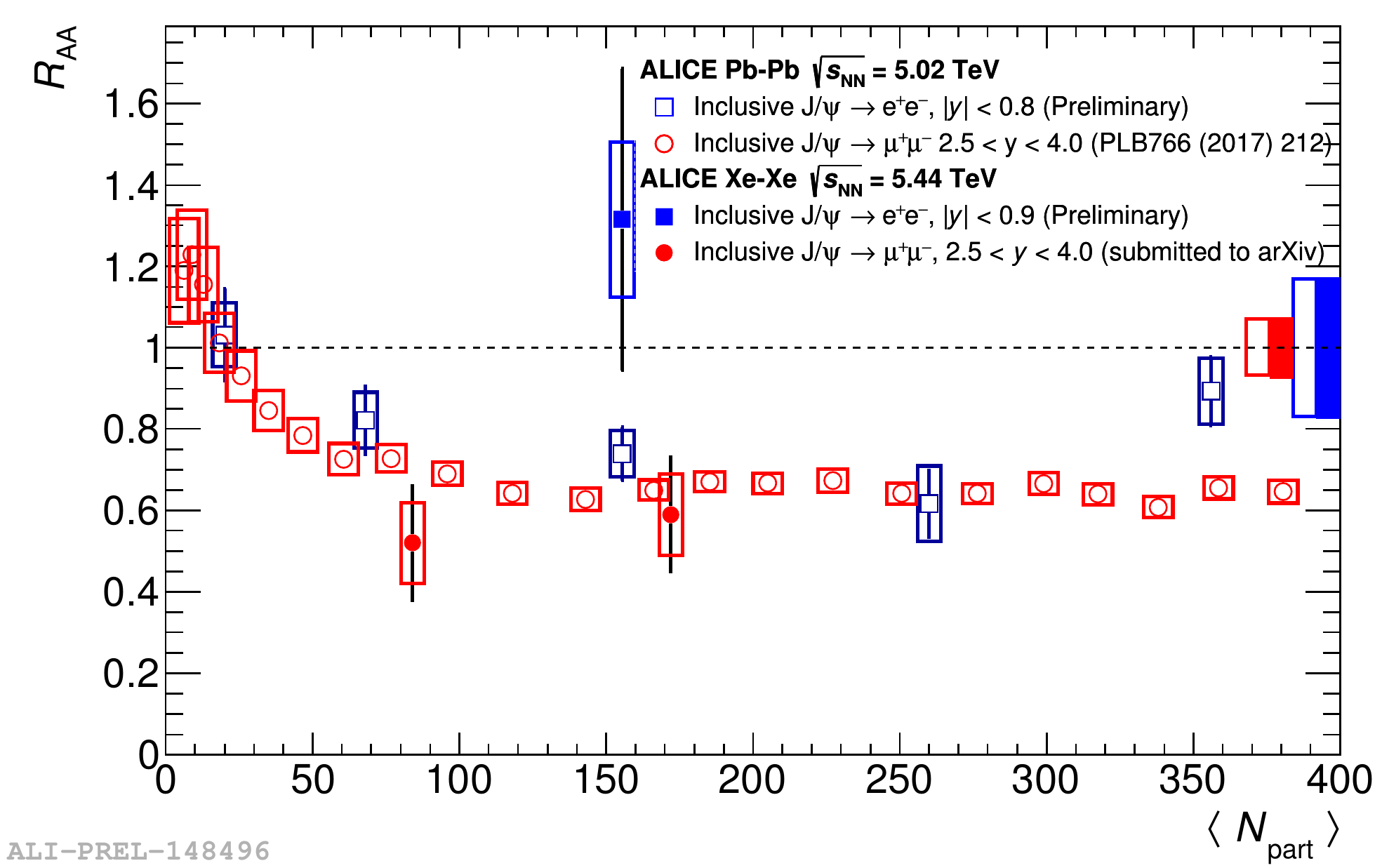}
	\caption{J/$\psi$ $R_{\mathrm{AA}}$ as a function of $\left\langle N_{\mathrm{part}} \right\rangle $. Red (open) circles: $2.5<y_{\mathrm{Lab}}<4$, $\sqrt{s_{\mathrm{NN}}} = 5.44~(5.02) ~\mathrm{TeV}$ Xe$-$Xe (Pb$-$Pb) collisions. Blue (open) squares: $0.9<|y_{\mathrm{Lab}}|$ ($0.8<|y_{\mathrm{Lab}}|$), $\sqrt{s_{\mathrm{NN}}} = 5.44~(5.02) ~\mathrm{TeV}$ Xe$-$Xe (Pb$-$Pb) collisions.}
	\label{fig:XeXeData}
	\vspace{0.5cm}
	\includegraphics[width=0.95\linewidth]{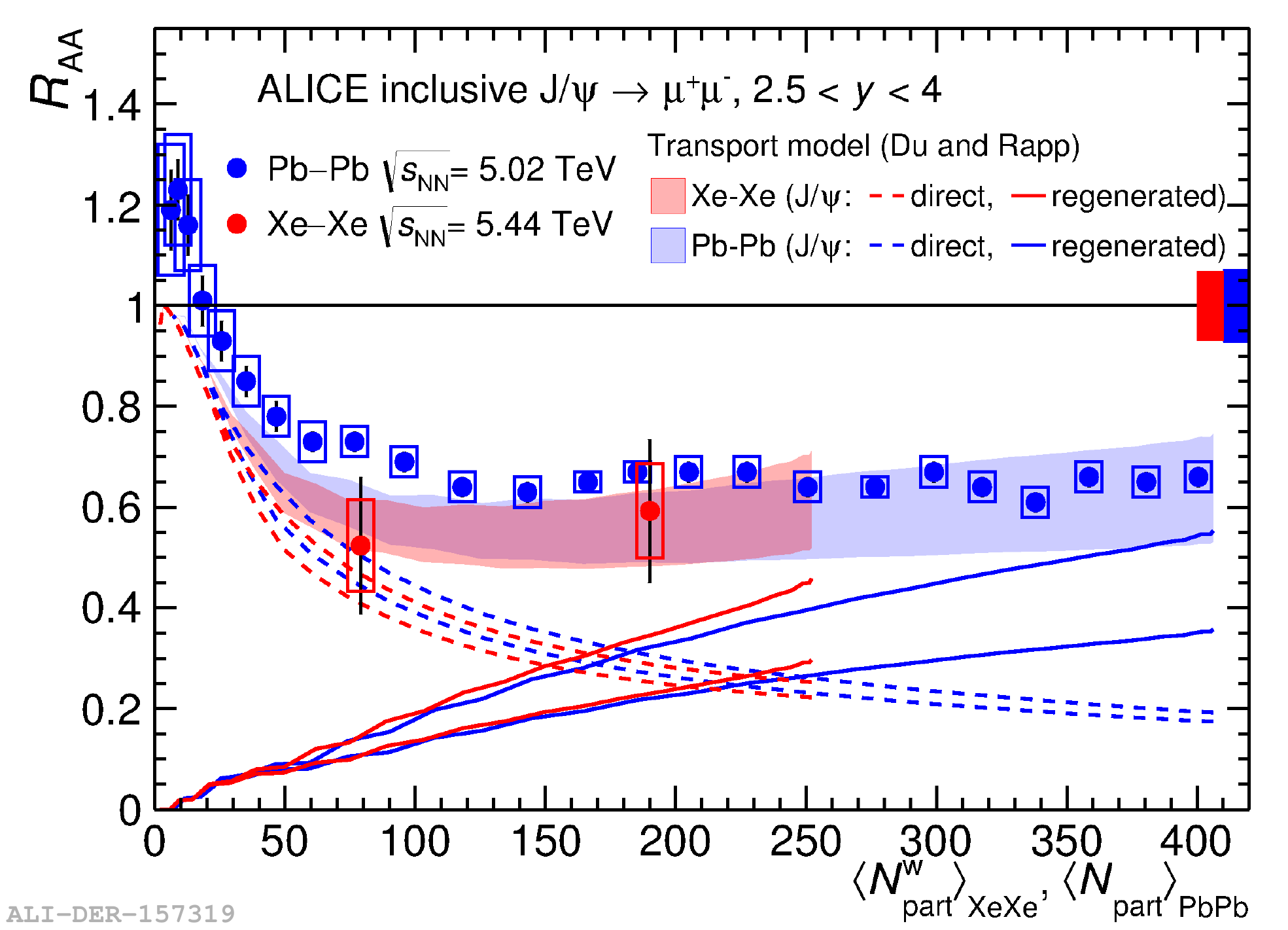}
	\caption{J/$\psi$ $R_{\mathrm{AA}}$ as a function of $\left\langle N_{\mathrm{part}} \right\rangle $ in the range of $2.5<y_{\mathrm{Lab}}<4$. Red (blue): $\sqrt{s_{\mathrm{NN}}} = 5.44~(5.02) ~\mathrm{TeV}$ Xe$-$Xe (Pb$-$Pb) collisions. Model calculations by Du and Rapp, dashed lines: direct, straight lines: regenerated, bands: incl. J/$\psi$. }
	\label{fig:XeXeFwdYDataTheory}
\end{wrapfigure}
\begin{figure}[h]
	\centering
	\subfloat[J/$\psi$ yield as a function of $y$, in the 20\% most central events in different $p_{\mathrm{T}}^{J/\psi}$ intervalls. \label{fig:MultiDiffYieldFwdY}]{\includegraphics[width=0.475\linewidth]{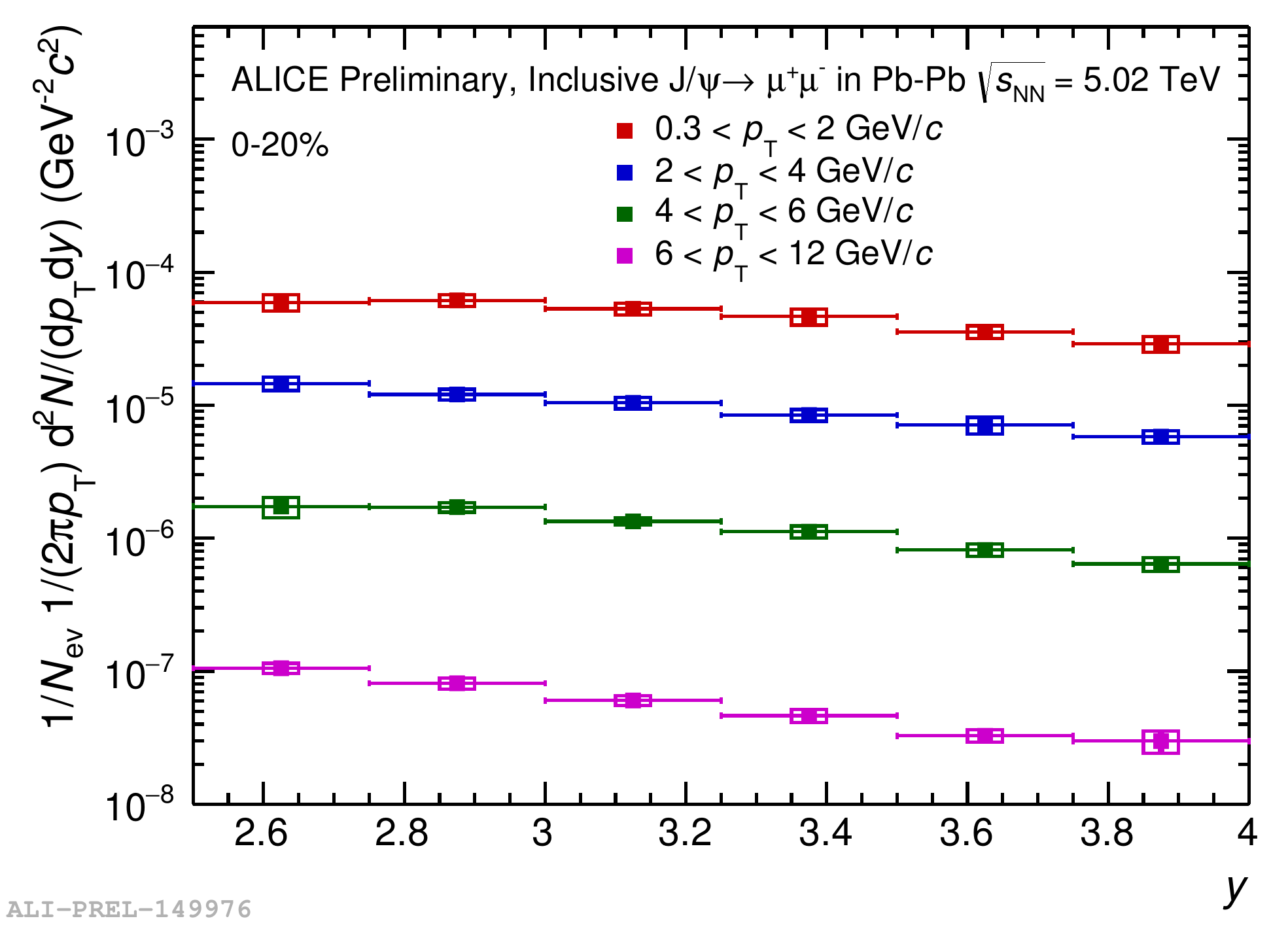}}
	\qquad
	\subfloat[Slope of an exponential fit to $\mathrm{d}^2N_{J/\psi}/\mathrm{d}p_{\mathrm{T}}\mathrm{d}y$ as a function of $p_{\mathrm{T}}^{J/\psi}$. \label{fig:MultiDiffSlopeFwdY}]{\includegraphics[width=0.475\linewidth]{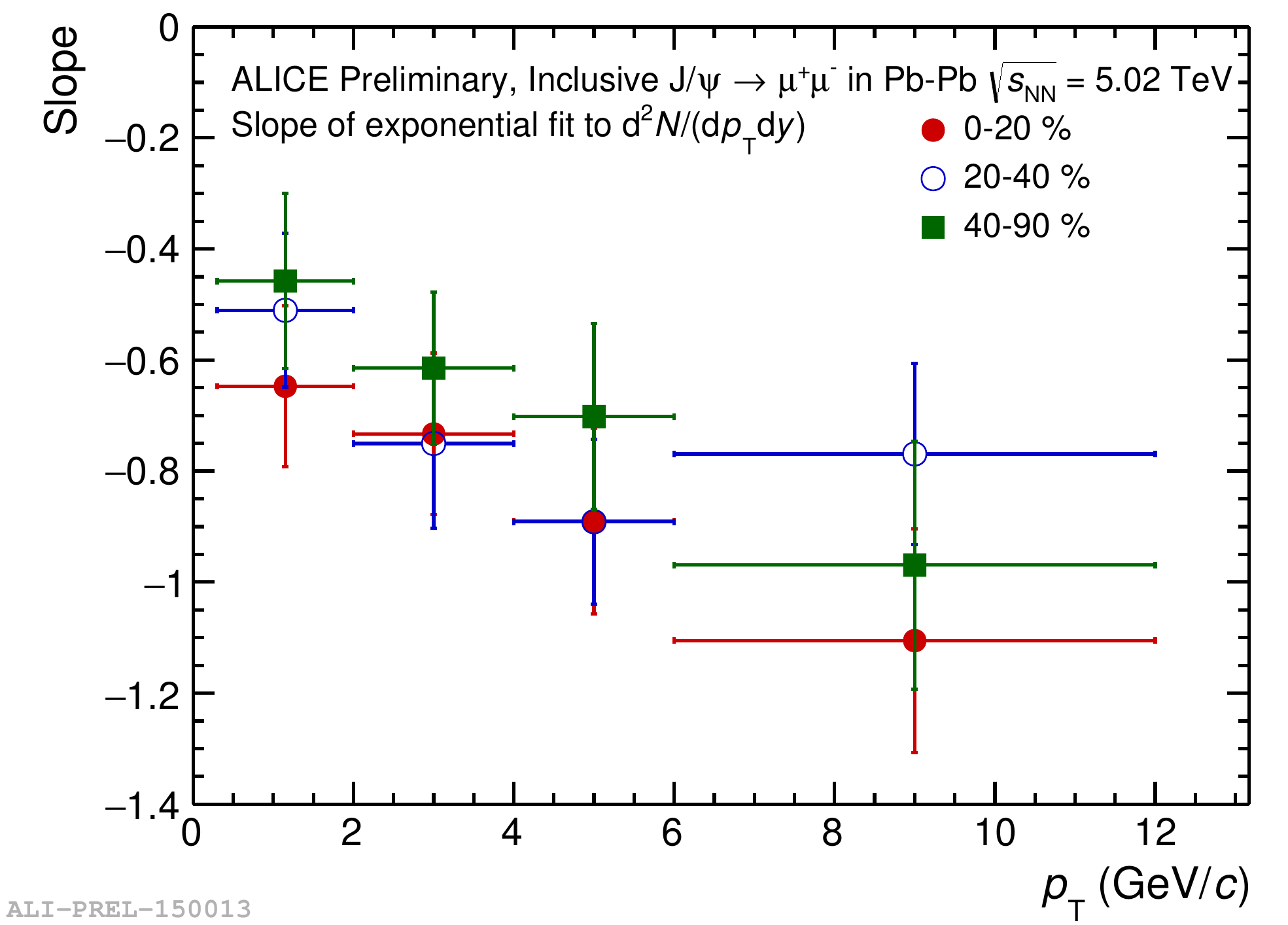}}%
	\caption{Preliminary results of a multi-differential J/$\psi$-yield analysis in $\sqrt{s_{\mathrm{NN}}} = 5.02 ~\mathrm{TeV}$ Pb$-$Pb collisions.}
	\label{fig:MultiDiffFwdY}
\end{figure}

\begin{figure}[h]
	\centering
	\begin{minipage}{.475\linewidth}
		\centering
		\includegraphics[width=.95\linewidth]{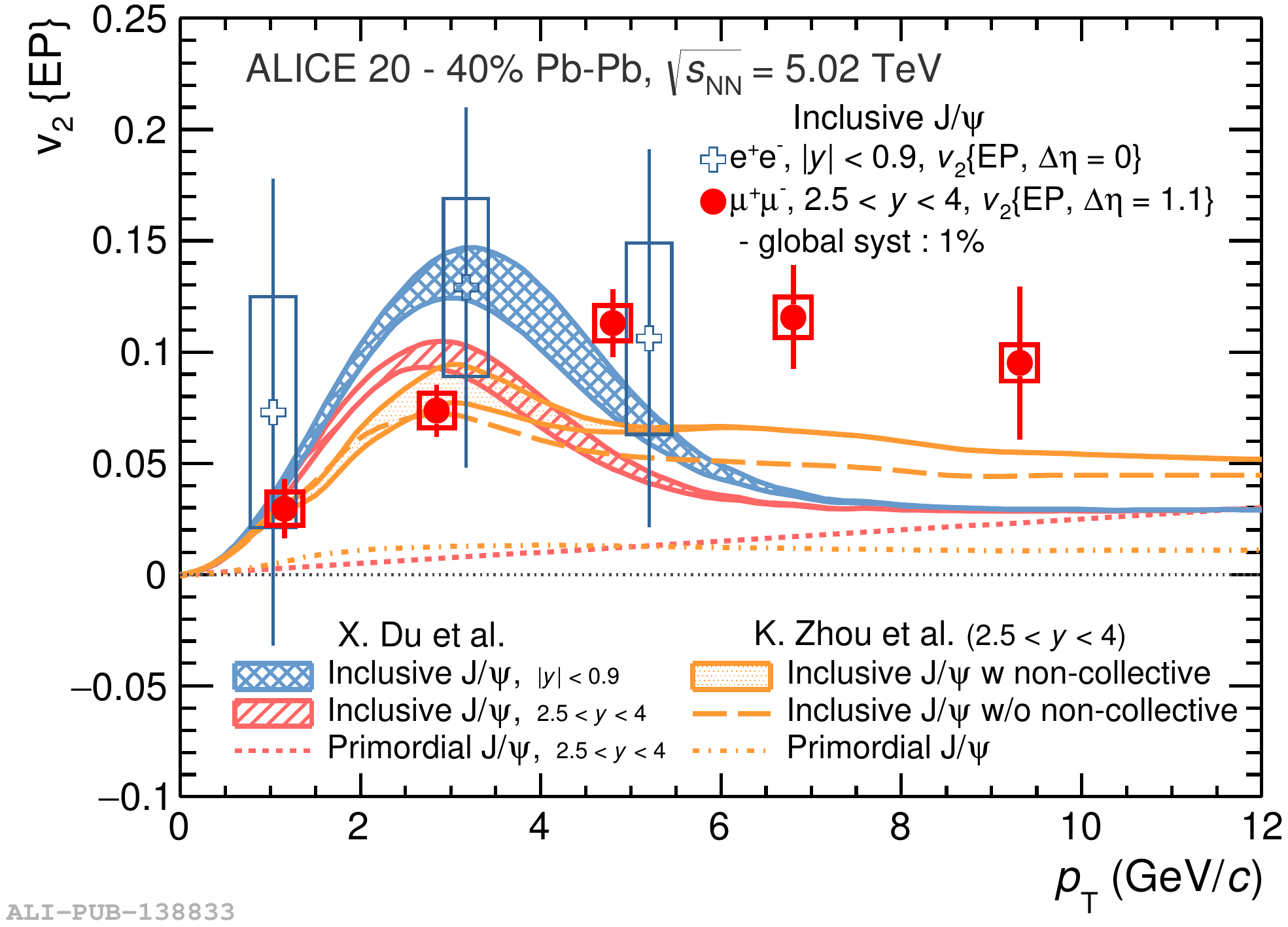}
		\caption{J/$\psi$ elliptic flow as a function of $p_{\mathrm{T}}$ at forward and mid-rapidity in semi-central $\sqrt{s_{\mathrm{NN}}} = 5.02 ~\mathrm{TeV}$ Pb$-$Pb collisions compared to transport model calculations \cite{Du:2015wha,Zhou:2014kka}.}
		\label{fig:EflowJpsiFwdMidYtheory}
	\end{minipage}%
	\qquad
	\begin{minipage}{.475\linewidth}
		\centering
		\includegraphics[width=.95\linewidth]{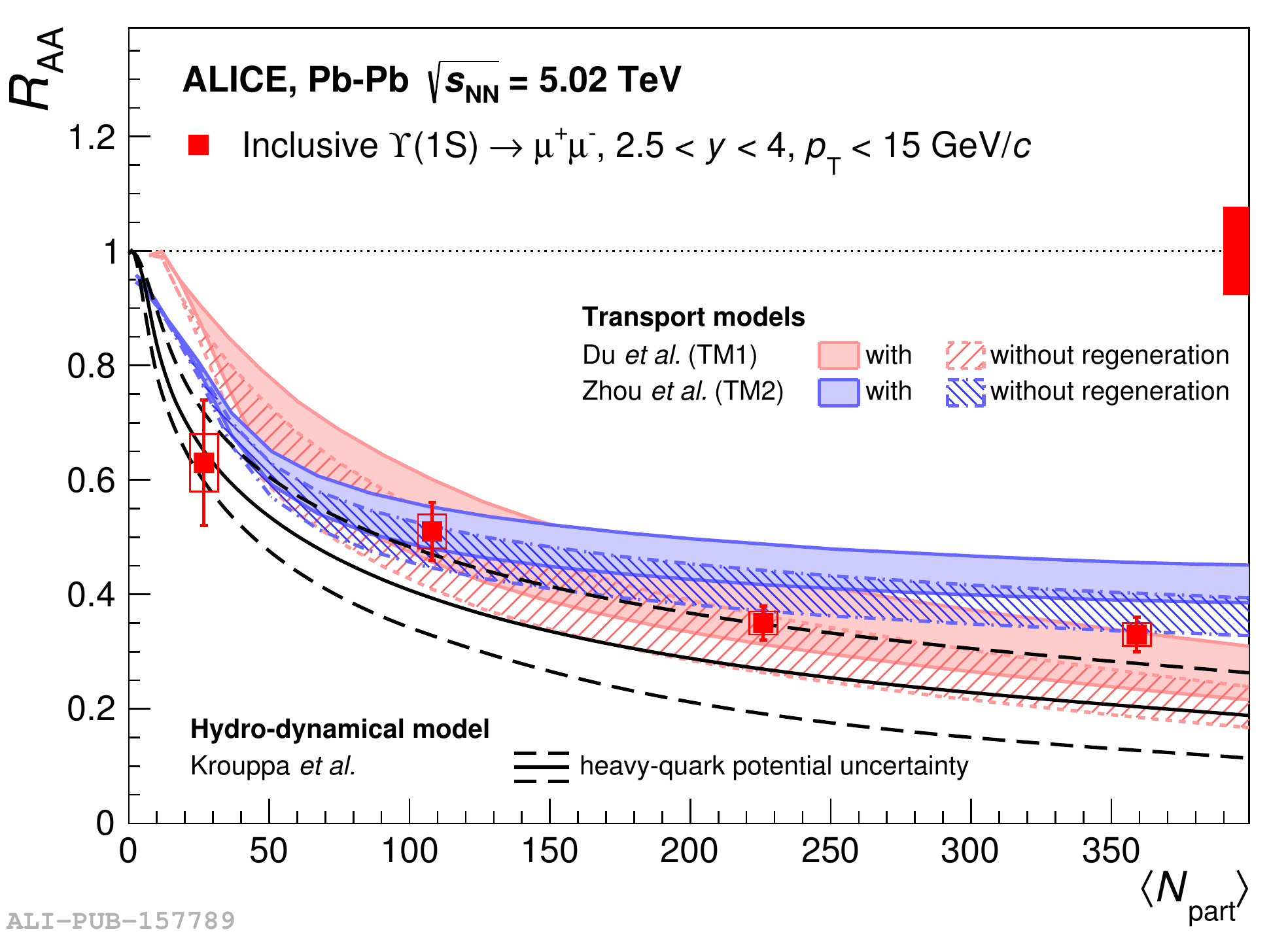}
		\caption{$\Upsilon$ $R_{\mathrm{AA}}$ as a function of $\left\langle N_{\mathrm{part}} \right\rangle $ in $\sqrt{s_{\mathrm{NN}}} = 5.02 ~\mathrm{TeV}$ Pb$-$Pb collisions, compared to transport \cite{Du:2017qkv,Zhou:2014hwa} and hydro-dynamical \cite{Krouppa:2017jlg} model calculations.}
		\label{fig:BottomoniumRaa}
	\end{minipage}
\end{figure}

 %
The ALICE experiment measures charmonia and bottomonia in their leptonic decay channels. The dimuon channel is measured at forward rapidity in the range of $2.5<y_{\mathrm{Lab}}<4$ with the muon spectrometer, while the dielectron channel is measured at mid-rapidity in the range of $-0.9<y_{\mathrm{Lab}}<0.9$ with the central-barrel detectors. A more detailed description of the ALICE experiment can be found in \cite{Aamodt:2008zz}. The results presented in the following are based on Pb$-$Pb ($A_{\mathrm{Pb}}=208$) and Xe$-$Xe ($A_{\mathrm{Xe}}=129$) collisions at $\sqrt{s_{\mathrm{NN}}} = 5.02~\mathrm{TeV}$ and $\sqrt{s_{\mathrm{NN}}} = 5.44 ~\mathrm{TeV}$, respectively, provided by the Large Hadron Collider (LHC). The ALICE experiment was capable to record integrated luminosities of $\mathcal{L}_{\mathrm{int}}^{\mathrm{Pb}} \approx 225~\mathrm{\mu b}^{-1}$ and $\mathcal{L}_{\mathrm{int}}^{\mathrm{Xe}} \approx 0.34~\mathrm{\mu b}^{-1}$ at forward rapidity and $\mathcal{L}_{\mathrm{int}}^{\mathrm{Pb}} \approx 13~\mathrm{\mu b}^{-1}$ and $\mathcal{L}_{\mathrm{int}}^{\mathrm{Xe}} \approx 0.25~\mathrm{\mu b}^{-1}$ at mid-rapidity, respectively. The large difference specially for the Pb$-$Pb sample is due the fact that at mid-rapidity minimum bias events were selected, while at forward rapidity special muon triggers were applied during data taking \cite{Adam:2016rdg,Acharya:2018jvc}.\\ \indent 
The J/$\psi$ $R_{\mathrm{AA}}$ measurement in Xe$-$Xe collisions at $\sqrt{s_{\mathrm{NN}}} = 5.44~\mathrm{TeV}$ enables the comparison between lighter (Xe) and heavier (Pb) nuclei. The measurement of the J/$\psi$ $R_{\mathrm{AA}}$ as a function of the average number of participants $\left( \left\langle N_{\mathrm{part}} \right\rangle \right)$ \cite{Abelev:2013qoq} in the forward \cite{Acharya:2018jvc}  and mid-rapidity range is shown in Fig. \ref{fig:XeXeData} together with the J/$\psi$ $R_{\mathrm{AA}}$ for Pb$-$Pb collisions. At forward rapidity the results for both collision systems agree with each other for similar $\left\langle N_{\mathrm{part}} \right\rangle $ within uncertainties. A comparison of the measurements at forward rapidity to a transport model \cite{Zhao:2011cv,Du:2015wha} is shown in Fig. \ref{fig:XeXeFwdYDataTheory}. The model is based on the thermal rate equation and contains continuous J/$\psi$ dissociation and regeneration in the QGP and the hadronic phase. The two collision systems as well as data and model agree well, which indicates that similar $\sqrt{s_{\mathrm{NN}}}$ and $\left\langle N_{\mathrm{part}} \right\rangle $ lead to similar relative contributions of suppression and (re)generation. However, strong conclusions are still difficult due to the uncertainties, which are for the model driven by the uncertainties on $\sigma_{c\bar{c}}$.\\ \indent
In J/$\psi$-$R_{\mathrm{AA}}$ measurements as a function of rapidity in $\sqrt{s_{\mathrm{NN}}} = 2.76 ~\mathrm{TeV}$ Pb$-$Pb collisions, ALICE observed an opposing trend between data \cite{Abelev:2013ila} and cold-nuclear-matter model calculations \cite{Vogt:2010aa,RAKOTOZAFINDRABE2011327}. These models expected an increase of $R_{\mathrm{AA}}$ with rapidity, while a decrease was observed in the data. The large data sample from $\sqrt{s_{\mathrm{NN}}} = 5.02 ~\mathrm{TeV}$ Pb$-$Pb collisions allows a multi-differential analysis of J/$\psi$ yields as a function of rapidity, transverse momentum and centrality \cite{PosterQMHushnud}, the preliminary results are shown in Fig. \ref{fig:MultiDiffFwdY}. The J/$\psi$ yield as a function of rapidity for different $p_{\mathrm{T}}$ intervals in the 20\% most central events is illustrated in Fig. \ref{fig:MultiDiffYieldFwdY}, the rapidity dependence is then fitted with exponential functions, the resulting slopes are depicted in Fig. \ref{fig:MultiDiffSlopeFwdY}. The slope decreases with increasing $p_{\mathrm{T}}$ and shows only a weak dependence on centrality.%
\\ \indent Another way to obtain more information about the different charmonium production mechanisms is the analysis of the elliptic flow. J/$\psi$ from (re)combined $c\bar{c}$ quarks should inherit the charm elliptic flow, which was observed in measurements of $D$-mesons \cite{Acharya:2017qps}. The results at forward rapidity show a significant positive J/$\psi$ elliptic flow in all studied $p_{\mathrm{T}}$ bins, the measurement at mid-rapidity agrees within large statistical uncertainties with the one at forward rapidity \cite{Acharya:2017tgv}. A comparison to model calculations \cite{Du:2015wha,Zhou:2014kka} is shown in Fig. \ref{fig:EflowJpsiFwdMidYtheory}. Both models contain a regeneration component, they describe the data in the low-$p_{\mathrm{T}}$ regime but clearly underestimate the elliptic flow at high $p_{\mathrm{T}}$. The question remains open how the elliptic flow of J/$\psi$ above $p_{\mathrm{T}} \approx 5$ GeV/$c$ is generated.
\\ \indent ALICE measured the $\Upsilon$(1S) $R_{\mathrm{AA}}$ as a function of $p_{\mathrm{T}}$ and $y$ as well as the inclusive $\Upsilon$(2S) $R_{\mathrm{AA}}$ in the forward rapidity range \cite{Acharya:2018mni}. An increase of the $\Upsilon$(1S) suppression towards more central events is observed, but due to the not precisely known feed-down fraction the amount of direct $\Upsilon$(1S) suppression is an open question. The measurement is shown in Fig. \ref{fig:BottomoniumRaa} and compared with three model calculations (two transport and one hydro-dynamical model calculation) \cite{Du:2017qkv,Zhou:2014hwa,Krouppa:2017jlg}, which all agree with the data within the uncertainties. The two transport models are shown without a regeneration component, both versions agree with the data, which indicates that a regeneration component for bottomonium should be negligible at LHC energies. The $\Upsilon$ (2S) suppression is significantly stronger then the one of the $\Upsilon$(1S) visible in the ratio of the $\Upsilon$ (2S) over $\Upsilon$(1S) $R_{\mathrm{AA}}$: $R_{\mathrm{AA}}^{\Upsilon (2\mathrm{S})}/R_{\mathrm{AA}}^{\Upsilon (1\mathrm{S})}=0.28 \pm 0.12(stat.) \pm 0.06(syst.)$. For the $\Upsilon$(1S) $R_{\mathrm{AA}}$ neither a significant dependence on $p_{\mathrm{T}}$ nor on $y$ is observed \cite{Acharya:2018mni}.
\section{Conclusions}
Measurements of J/$\psi$ $R_{\mathrm{AA}}$ in Xe$-$Xe collisions at $\sqrt{s_{\mathrm{NN}}} = 5.44 ~\mathrm{TeV}$, the elliptic flow and multi-differential yields of J/$\psi$ as well as the differential (inclusive) $\Upsilon$(1S) ($\Upsilon$(2S)) $R_{\mathrm{AA}}$ in Pb$-$Pb collisions at $\sqrt{s_{\mathrm{NN}}} = 5.02~\mathrm{TeV}$ have been presented. The results indicate that quarkonia production at the LHC is a combination of suppression and (re)generation, strongly dependent on the $\sigma_{Q\bar{Q}}$. However, there are still unanswered questions, e.g. the reason for the significant J/$\psi$ elliptic flow at higher $p_{\mathrm{T}}$ or the amount of direct $\Upsilon$(1S) suppression, which hopefully can be answered in the near future.%



\bibliographystyle{elsarticle-num}
\bibliography{qm2018ProceedingsPD}







\end{document}